\documentclass[aps,pre,preprint,groupedaddress]{revtex4}

\usepackage{graphicx}
\usepackage{latexsym}

\begin{document}

%\preprint{}

\title{
Folding of the triangular lattice in a discrete three-dimensional space:
Crumpling transitions in the negative-bending-rigidity regime
}

\author{Yoshihiro Nishiyama}
%\email[]{Your e-mail address}
%\homepage[]{Your web page}
%\thanks{}
%\altaffiliation{}
\affiliation{Department of Physics, Faculty of Science,
Okayama University, Okayama 700-8530, Japan.}

\date{\today}

\begin{abstract}
Folding of the triangular lattice in a discrete three-dimensional space
is studied numerically.
Such ``discrete folding''
was introduced by Bowick and co-workers as a simplified version 
of the polymerized membrane in thermal equilibrium.
According to their cluster-variation method (CVM) analysis,
there appear various types of phases as the bending rigidity $K$ changes 
in the range $-\infty < K < \infty$.
In this paper, 
we investigate the $K<0$ regime, for which the
CVM analysis with the single-hexagon-cluster approximation predicts two 
types of (crumpling) transitions of 
both continuous and discontinuous characters.
We diagonalized the transfer matrix for the strip widths up to $L=26$
with the aid of the
density-matrix renormalization group.
Thereby, we found that discontinuous transitions occur successively 
at $K=-0.76(1)$ and $-0.32(1)$.
Actually, these transitions are accompanied with distinct hysteresis effects.
On the contrary, the latent-heat releases are
suppressed considerably as $Q=0.03(2)$ and $0.04(2)$ for respective transitions.
These results indicate that the
singularity of crumpling transition
can turn into a {\em weak}-first-order type by 
appreciating the fluctuations beyond a meanfield level.
\end{abstract}

% insert suggested PACS numbers in braces on next line
% insert suggested keywords - APS authors don't need to do this
%\keywords{}

%\maketitle must follow title, authors, abstract, \pacs, and \keywords
\maketitle

\section{\label{section1}Introduction}

A model of
``discrete folding'' was introduced 
as a possible lattice realization (discretized version) for
the tethered (polymerized) membranes in thermal equilibrium
\cite{Kantor90,DiFrancesco94a,DiFrancesco94b,Cirillo96b,%
Bowick95,Cirillo96,Bowick96,Bowick97}.
The constituent molecules of the tethered membranes
are connected with each other via polymerization
\cite{Nelson89,Nelson96,Bowick01}.
Hence, the in-plane strain is subjected to finite shear moduli.
In this sense, the tethered membranes are stiff,
compared with the fluid membranes \cite{Canham70,Helfrich73} for which
the constituent molecules are diffusive, and cannot support a shear.
Reflecting this peculiarity,
the tethered membranes get flattened macroscopically
at sufficiently low temperatures 
(or equivalently large bending rigidities) \cite{Nelson87}.
This phase transition is called the crumpling transition.
(Because the crumpling transition occurs even in the absence of the
self-avoidance, we neglect this effect throughout this paper.)
The flat phase is characterized by the 
long-range orientational order
of the surface normals.
It would be remarkable that such
an $O(3)$ symmetry, namely, an orientational order,
breaks spontaneously for a two-dimensional manifold.
Actually, it has been known that the fluid membranes
are always crumpled irrespective of the temperatures.
To clarify the nature of the crumpling transition of the tethered membranes,
a good deal of theoretical analyses have been reported so far.
However, there still remain controversies whether the singularity
belongs to a continuous phase transition 
\cite{Kantor86,Kantor87,Baig89,Ambjorn89,Renken96,Harnish96,%
Baig94,Bowick96b,Wheater93,Wheater96,David88,Doussal92,Espriu96}
or a discontinuous
one accompanied with appreciable latent-heat release
\cite{Paczuski88,Kownacki02}.
Actually, in numerical simulations, it is not quite obvious to rule out a possibility
of {\em weak}-first-order phase transition \cite{Kantor87b,Kantor87c}.

As would be anticipated, 
an approach via the discrete-folding model 
\cite{Bowick95,Cirillo96,Bowick96,Bowick97}
is promising to resolve this longstanding issue:
In fact, the discrete-folding model
reduces to a mere Ising magnet \cite{Bowick95}
with a dual transformation; we explain the details afterward.
The Ising variables, however, are subjected to 
a constraint, which makes the thermodynamics rather nontrivial.
Nevertheless, once formulating the discrete folding
in terms of the Ising variable,
we are able to resort to a variety of techniques developed 
in the study of magnetism:
Cirillo and co-workers carried out an extensive
cluster-variation method (CVM) analysis with the single-hexagon-cluster
approximation \cite{Cirillo96,Bowick97}.
Remarkably enough, according to their CVM solution,
various types of phases emerge as the bending rigidity $K$ changes;
see Fig. \ref{figure1} for a schematic drawing of the phase diagram.
As depicted in the figure, the symbols, 
$K_{P3}$, $K_{G3}$, and $K_{C3}$, specify the respective
crumpling-transition points separating
the piled-up, tetrahedral,  octahedral, and totally-flat phases;
we followed the notation in Ref. \cite{Bowick97}.

Stimulated by their result, in an earlier paper \cite{Nishiyama04},
we performed a first-principles simulation for the discrete folding.
Our simulation scheme is based on the transfer-matrix diagonalization
with the aid of the density-matrix renormalization group
\cite{White92,White93,Nishino95,Nishiyama02,Nishiyama03}.
In Ref. \cite{Nishiyama04}, we dwelt on the $K>0$ regime,
and investigated the crumpling transition at $K=K_{C3}$:
We estimated the transition point as $K_{C3}=0.195(2)$, and 
estimated the amount of the latent heat as $Q=0.365(5)$.
Moreover, we confirmed that
for $K>K_{C3}$,
the entropy disappears completely as predicted by the CVM analysis 
\cite{Cirillo96}.
That is, the membrane stretches out into a totally flat sheet
in that regime.
In this sense, the transition at $K=K_{C3}$ is rather peculiar,
and it would be less relevant to that of the realistic tethered membranes. 
Beside slight quantitative discrepancies concerning 
the folding entropy at $K=0$ and the amount of the latent heat,
our simulation data support the CVM predictions,
implying that the membrane undulations beyond a
single-hexagon cluster are not quite significant in the $K>0$ regime.

In this paper, we devote ourselves to the $K<0$ regime, where the CVM analysis
predicts two types of crumpling transitions \cite{Bowick97}.
Our aim is to examine the singularities of these transitions
with the density-matrix renormalization group for $L \le 26$:
It is conceivable that the order of the transition changes or that 
even the transition disappears
by appreciating the fluctuations beyond the single-hexagon-cluster level.
As a consequence,
we found that the crumpling transitions occur
at $K=-0.76(1)$ and $-0.32(1)$, and the singularities are both discontinuous.
Actually, we estimated the amounts of the latent heat as $Q=0.03(2)$ and $0.04(2)$, respectively.
Our result indicates that the singularity of the crumpling transition
can change into a (weak) first-order type by fully appreciating the fluctuations.

In fairness, it has to be mentioned that
the discrete folding has been studied extensively with computer simulations
other than the density-matrix renormalization group,
namely,
the conventional full-diagonalization scheme 
\cite{Bowick96,Bowick97} and the Monte Carlo method
 \cite{Pujol04}:
In the full-diagonalization calculation,
the tractable systems sizes are limited within $L \le 6$,
and are not feasible for identifying the characters  
of the transitions.
In fact, as for the transition at $K=K_{P3}$, 
it was speculated that no sign of the transition 
could be captured until $L=8$, which
is far beyond the ability of the full-diagonalization scheme.
On the other hand, the Monte Carlo simulation
suffers from a very slow relaxation to the thermal equilibrium
(glassy behavior).
Although such a metastability is a fascinating topic in its own right,
it brings about a severe obstacle to an efficient sampling in thermal equilibrium.
(Such slow relaxation is to be attributed to
the constraint to the Ising variables.)
We stress that the density-matrix renormalization group
is capable of treating large system sizes, and free from the slow-relaxation 
problem.

The rest of this paper is organized as follows.
In the following section, we give an account of the discrete folding,
following the presentation in Ref. \cite{Bowick95}.
Then, we explain the density-matrix renormalization group adopted to
the discrete-folding model \cite{Nishiyama04}.
In Sec. \ref{section3}, we present the numerical results.
The last section is devoted to the summary and discussions.

\section{\label{section2}
Diagonalization of the transfer matrix with
the density-matrix renormalization group:
A brief reminder of Ref. \cite{Nishiyama04}
}

In this section, we explain an outline of our simulation scheme.
Full account of details will be found in an earlier paper \cite{Nishiyama04}.
To begin with, we introduce the discrete folding of the triangular lattice in 
a discretized three-dimensional space \cite{Bowick95}.

\subsection{Discrete folding in three dimensions \cite{Bowick95}}

Let us explain how the triangular lattice is
folded (embedded) in a discrete three-dimensional space, namely, 
the face-centered cubic (fcc) lattice.
As shown in Fig. 3 of Ref. \cite{Bowick95},
the fcc lattice is viewed as a close packing of two types of polygons,
namely, the 
octahedrons and tetrahedrons,
whose vertexes are shearing the fcc-lattice points;
the faces of these polygons are all equilateral triangles.
Hence, wrapping these polygons with a sheet of the triangular lattice,
we are able to embed the sheet
arbitrarily in the fcc lattice.
More specifically,
the relative fold angles $\theta$ between adjacent triangles
are taken from four possibilities, namely,
``no fold'' ($\theta=\pi$),
``complete fold'' ($\theta=0$)
``acute fold'' ($\theta= acos (1/3)$), and ``obtuse fold'' 
($\theta = acos (-1/3)$).

The above discretization leads to an Ising-spin representation 
for the discrete folding.
An efficient representation, the so-called gauge rule, 
reads as follows \cite{Bowick95}:
We place two types of Ising variables,
namely, $\sigma_i$ and $z_i$,
at each triangle (rather than each joint); see Fig. \ref{figure2} (a).
The gauge rule states that these Ising spins specify the joint angle
between the adjacent triangles.
That is, provided that the $z$ spins are antiparallel ($z_1 z_2=-1$)
for a pair of adjacent neighbors, 
the joint angle is either acute or obtuse fold.
Similarly, if $\sigma_1 \sigma_2=-1$ holds, 
the relative angle is either complete or obtuse fold.
Note that the above rules specify the joint angle unambiguously.
To summarize,
we introduced Ising variables placed at each triangle.
Because the structure dual to the triangular lattice is hexagonal,
the Ising variables constitute an Ising model on the hexagon lattice.
Hence, the transfer-matrix strip looks like that drawn 
in Fig. \ref{figure2} (b).
The row-to-row statistical weight $T_{\{\sigma_i,z_i\},\{\sigma_i',z_i'\}}$
yields the transfer-matrix element.
However, according to Ref. \cite{Bowick95}, the
Ising variables are not quite independent, but subjected to some
constraints for each hexagon.
The transfer-matrix element is given by the following form
with extra factors that enforce the constraints \cite{Bowick95};
\begin{equation}
\label{transfer_matrix}
T_{ \{z,\sigma\},\{z',\sigma'\} } = 
    ( \prod_{j=1}^{L-1} U_j V_j ) \exp( -H/T )      ,
\end{equation}
with,
\begin{equation}
\label{constraint1}
U_j= \delta(
 \sigma_{2j-2}+\sigma_{2j-1}+\sigma_{2j}
    +\sigma'_{2j-1}+\sigma'_{2j}+\sigma'_{2j+1}\ mod\ 3,0)   ,
\end{equation}
and,
\begin{equation}
\label{constraint2}
V_j=  \prod_{c=1}^{2} \delta(
\alpha_c(z_{2j},z_{2j-1},z_{2j-2},z'_{2j-1},z'_{2j},z'_{2j+1})\ mod\ 2,0)
                                    .
\end{equation}
Here, $\delta(m,n)$ denotes Kronecker's symbol, and $\alpha_c$
is given by,
\begin{equation}
\alpha_c(z_1, \dots ,z_6)=\sum_{i=1}^6
\frac{1}{2}(1-z_i z_{i+1}) \delta(c_0+\sum_{j=1}^i \sigma_i - c\ mod\ 3,0)  .
\end{equation}
%(These constraints, Eqs. (\ref{constraint1}) and (\ref{constraint2}),
%allow 96
%spin configurations around each hexagon \cite{Bowick95}.
%In that sense,
%the resultant model is regarded as a 96-vertex model.)
The Boltzmann factor $\exp( -H/T )$ in Eq. (\ref{transfer_matrix}) is due to
the bending-energy cost. 
Hereafter, we choose the temperature $T$ as the unit of energy; 
namely, we set $T=1$.
As usual,
the bending energy is given by the inner product, $\cos \theta_{ij}$ of the surface
normals of adjacent triangles.
Hence, the bending energy is given by the formula;
\begin{equation}
\label{elastic_energy}
H=  - 0.5  \sum_{\langle i j \rangle} 
 K \cos \theta_{ij} = - 0.5 \sum_{\langle i j \rangle}
              \frac{1}{3}K \sigma_i \sigma_j (1+2 z_i z_j)  ,
\end{equation}
with the bending rigidity $K$.
Here, the summation $\sum_{\langle ij \rangle}$ runs over all possible 
nearest-neighbor
pairs $\langle ij \rangle$ around each hexagon.  
(The overall factor $0.5$ is intended to reconcile the double counting.)
The above completes a prescription to construct the transfer matrix.
In the following, we explain how we diagonalized the transfer matrix.

\subsection{Diagonalization of the
transfer matrix for the discrete folding:
An application of the density-matrix renormalization group \cite{Nishiyama04}}

As would be apparent from Fig. \ref{figure2} (b),
a unit cell contains four Ising variables.
Therefore,
the size of the transfer matrix grows rapidly in the form 
$16^L \times 16^L$ with
the system size $L$.
% the definition of $L$ is shown in Fig. \ref{figure2} (b).
Hence, the diagonalization of the transfer matrix requires
huge computer-memory space.
(Actually, with the conventional full-diagonalization scheme,
the tractable systems sizes are limited within 
$L \le 6$ \cite{Bowick95,Bowick96,Bowick97}.)
In order to cope with this difficulty, in an earlier paper \cite{Nishiyama04},
we developed an alternative (memory saving) diagonalization method
based on the density-matrix renormalization group.
\cite{White92,White93,Nishino95}.
In the following, we outline the algorithm,
placing an emphasis on the changes specific to this problem;
refer to a textbook \cite{Peschel99} for a pedagogical guide.

%Moreover, the open boundary condition must be imposed
%in order to release the constraints from the boundaries;
%otherwise, the constraints are too restrictive.
%The boundary effect deteriorates the data so that the data
%acquire severe finite-size corrections.
%In the following section, we propose an alternative diagonalization
%scheme which enables us to treat large system sizes.

%The method allows us to treat large system sizes.
%Our algorithm is standard, and we refer the reader to consult with
%the text \cite{Peschel99} for technical details.

The density-matrix renormalization group is based on the idea of
the (computer aided) real-space decimation.
We presented a schematic drawing of 
one operation of the real-space-decimation procedure in Fig. \ref{figure3}.
Through the decimation, the block states and the adjacent spin variables
(hexagon)
are renormalized altogether into a new block states.
Note that through the decimation, the number of states for the
(renormalized) block is truncated within $m$;
the parameter $m$ sets the simulation precision.
Hence, sequential applications of the procedure enable us to reach
very long system sizes.
In order to reduce the truncation error, we need to retain 
$m$ significant bases.
According to the criterion advocated in Ref. \cite{White92,White93},
such $m$ states are chosen from the eigenstates
of the 
(local) density matrix 
with dominant statistical weights (large eigenvalues) $\{ w_\alpha \}$
($\alpha$: integer index)
\cite{Peschel99}.
% see Ref. \cite{Peschel99} for a pedagogical guide.
% This is an essential part of the density-matrix renormalization group 
% \cite{White92}, and the name comes from this.

This is a good position to address a few remarks regarding
the changes specific to the present problem.
First, in our simulation, in order to reduce the truncation error
of the real-space decimation,
we adopted the ``finite-size method'' \cite{White93,Peschel99}.
%We managed 3 sweeps at least, and confirmed that a good convergence
%is achieved
%in the sense that the error is negligible compared with the finite-size corrections.
Secondly, we applied a magnetic field $h$ 
at an end of the transfer-matrix strip.
Namely,
we incorporated an additional Hamiltonian of the form
$-h \delta_{\sigma,1}\delta_{z,1}$ with $h=0.05$ and
 the edge spins $(\sigma,z)$.
This trick was utilized in the preceding full-diagonalization
calculation \cite{Bowick97},
and it aims to split off the degeneracy caused by the
trivial four-fold symmetry ($\sigma=\pm1,z=\pm1$) of the overall 
membrane orientation (gauge redundancy).

\section{\label{section3}Numerical results}

In this section, 
based on the algorithm explained above,
we perform computer simulations for the discrete-folding model 
in the negative
bending rigidity ($K<0$) regime.
As a preliminary survey,
we calculate the order parameters, Eqs. 
(\ref{order1})-(\ref{order2}), for a wide range of $K$.
Thereby, we examine
the phase diagram predicted by
the CVM analysis as depicted in Fig. \ref{figure1}.
We will also provide a distribution of the density-matrix eigenvalues
in order to elucidate the reliability of the present scheme.
These preliminary analyses are followed by
large-scale simulations to determine the singularities of the 
crumpling transitions
precisely.

\subsection{Preliminary survey: 
A check of the reliability of the simulation scheme}

In Fig. \ref{figure4}, we plotted the following order parameters (local magnetizations)
\cite{Bowick97}
for the system size $L=14$ and the number of states remained for 
a renormalized block $m=12$;
\begin{eqnarray}
\label{order1}
O  &=& \langle \sigma_s \rangle \\
\label{order2}
T &=& \langle z \sigma_s \rangle \\
\label{order3}
P &=& \langle z \rangle  ,
\end{eqnarray}
where the bracket $\langle \cdots \rangle$ denotes the thermal average,
and the symbol $\sigma_s$ stands for the staggered component of the moment $\sigma$.
(Non-zero magnetizations are induced 
by the symmetry-breaking field at an end;
see Sec. \ref{section2} for details.)
As depicted in Fig. \ref{figure1}, the CVM analysis predicts successive crumpling transitions
with varying the bending rigidity $K$, and these phases are characterized by the above
order parameters.
Indeed, our simulation data support this claim;
namely,
there appear crumpling transitions at $K\approx -0.9$ and $-0.3$ successively.
However, on closer inspection, our simulation data exhibit clear
evidences of the hysteresis 
effects for both transitions.
Hence, contrary to the CVM scenario, both transitions should be of discontinuous
characters.

Before going into detailed analysis of the hysteresis effects,
let us check the reliability of the scheme.
In Fig. \ref{figure5}, we show the distribution of 
the statistical weights
(eigenvalues of the density matrix)
$\{ w_\alpha \}$ for $K=-0.2$, $L=15$ and $m=15$.
We see that the statistical weight drops very rapidly.
In other words, a majority of the statistical weight
concentrates on a few dominant (relevant) states.

From the data,
we are able to read off a reliability of the numerical simulation:
Because we discard those states with the indexes
$\alpha>m$ through an operation of the renormalization-group
procedure,
the statistical weight $w_{\alpha}$ at $\alpha=m$ 
yields an indicator of the truncation error.
To be specific, the fifteenth state, for instance, 
exhibits a very tiny statistical 
weight $w_{15} \approx 5 \times 10^{-4}$.
%%%That is, remaining only fifteen significant states
%%%for a renormalized block,
%%%we attain a precision of the order $\sim 5 \times 10^{-4}$.
%%%% In this paper, we retain, at most, $m=...$ bases for a renormalized block.
%%%%%%%%%%%%%%%%
This truncation error accounts for a precision of the free energy.
The reliability of 
other quantities such as the internal energy $E$ 
could be enhanced by a factor of 10; namely, the precision would be of the order $ \sim 5 \times 10^{-3}$.
The data scatter of Figs. \ref{figure8} and \ref{figure10}, for instance,
should be attributed to this uncertainty.
%%%%%
%This truncation error
%The truncation error is almost negligible in a practical sense,
%and actually,
%it is smaller than the finite-size corrections.
Moreover, it is to be noted that the strength of 
bending rigidity set in Fig. \ref{figure5},
namely, $|K|=0.2$,
is rather small.
Therefore, upon stiffening the rigidity $|K|$ further, the fluctuations get suppressed,
and correspondingly, the truncation error improves.
In particular, for $|K| > 0.7 (\approx |K_{P3}|) $, such an improvement 
appears to be pronounced.
Encouraged by these findings,
we proceed to large-scale simulations to determine 
the singularities of the crumpling transitions.

\subsection{Singularity of the crumpling transition at  $K=K_{G3}$}

In the above, we observed clear evidences of the onsets of the crumpling transitions
at $K \approx -0.9$ and $-0.3$.
Contrary to the CVM scenario, these transitions are both discontinuous.
Here, performing large-scale simulations, we confirm that the transitions
are indeed discontinuous ones accompanied with appreciable latent-heat releases.

We first consider the crumpling transition at $K=K_{G3}$,
namely, the transition separating the tetrahedral and octahedral phases.
In Fig. \ref{figure6}, we plotted the free energy per triangle
$f_{\rightarrow,\leftarrow}$
for $L=20$, and $m=15$.
(See our preceding paper \cite{Nishiyama04} for an algorithm 
to calculate the free energy reliably.)
As indicated in the plot, we swept the parameter $K$
in both descending and ascending directions.
Correspondingly, the indexes of
$f_\leftarrow$ and $f_\rightarrow$ specify the parameter-sweep directions.
From the data, we observe a clear hysteresis effect;
the intersection point of $f_\rightarrow$ and $f_\leftarrow$ 
yields the
location of the first-order phase transition.
Therefore, we confirm that the crumpling transition at $K=K_{G3}$ 
is indeed a first-order phase transition.

In order to determine the transition point more precisely,
in Fig. \ref{figure7},
we plotted the excess free energy $f_\leftarrow-f_\rightarrow$ 
for various values of $L$ and $m$.
From the figure, we see that the data for $L \ge 20$
converge to the thermodynamic limit satisfactorily.
From the coexistence condition $f_\leftarrow-f_\rightarrow=0$,
we estimate the transition point as $K=-0.32(1)$.
The location of the transition point is in good agreement
with the CVM result $K=-0.294$.

In order to estimate the latent heat,
in Fig. \ref{figure8}, 
we plotted the residual internal energy $E_\leftarrow-E_\rightarrow$ per triangle
for various values of $L$ and $m$.
Similar to the above, the data, $E_{\rightarrow}$ and $E_{\leftarrow}$,
are calculated by
sweeping $K$ in ascending and descending directions, respectively.
The residual internal energy at the transition point $K=-0.32(1)$
yields the latent heat.
The data for $L \ge 20$ appear to reach the thermodynamic limit satisfactorily.
From these data, we determine the latent heat as $Q=0.04(2)$.
The present estimate is not consistent with the CVM result
$Q=0.14$;
this value is taken from
Fig. 12 of Ref. \cite{Bowick97}.
We see that the discrepancy between the present estimate and
the CVM analysis with a single-hexagon-cluster approximation 
is rather severe, and
it may indicate that
the fluctuations beyond a hexagon cluster are not quite negligible.
Such fluctuations contribute to smearing out the
latent-heat release significantly.

Although the latent-heat release acquires such a notable
suppression,
there appears a distinct hysteresis effect as shown in Figs. 
\ref{figure4} and \ref{figure6}.
This peculiarity may be reflected in the preceding
full-diagonalization calculation for $L \le 6$
\cite{Bowick97}:
This calculation 
captures an evidence of the onset of the transition 
by a steep increase of the order parameter $T$, whereas
the character of the singularity was remained unsolved.
Actually, such a small amount of latent heat could hardly be resolved 
by conventional approaches.
Nevertheless, such peculiarities, namely, a small latent heat and  
a distinct hysteresis effect, suggest that
the barrier between the coexisting states
is hardly surmountable,
albeit these states are rather close to each other in the configuration space,
To the best of our knowledge, such a feature
is specific to the crumpling transition, and it may be related to
the anomalous metastability (glassy behavior)
observed in the Monte Carlo simulation
\cite{Pujol04}.

Lastly, let us address a technical remark on the hysteresis effect.
The hysteresis cannot be reproduced by the conventional
full-diagonalization scheme.
In the density-matrix renormalization group, on the contrary,
in the course of simulation,
past information
is encoded (retained) in the renormalized block
through the sequential applications of normalizations.
Because of this ``memory effect'' \cite{Gendiar02}, 
the density-matrix renormalization group reproduces a hysteresis behavior.
Beside this advantage,
the method admits reliable estimate for the free energy.
These advantages are significant to determine the 
first-order-phase-transition point reliably.

\subsection{Singularity of the crumpling transition at $K=K_{P3}$}

Let us turn to surveying the singularity at $K=K_{P3}$, namely,
the crumpling transition separating the piled-up and tetragonal phases;
note that according to CVM, the transition should be continuous.

In Fig. \ref{figure9}, we plotted the excess free energy 
$f_\leftarrow-f_\rightarrow$ 
for various values of $L$ and $m$.
The data exhibit a clear hysteresis effect.
Hence, we confirm that contrary to the CVM scenario,
the transition is of a discontinuous character.
From the coexistence condition $f_\leftarrow-f_\rightarrow=0$, we determine the
transition point as $K=-0.76(1)$.
The location of the transition point is
consistent with the CVM estimate $K=-0.852$.
Hence, apart from the singularity characterization,
the CVM treatment appears to be justified.

%%%%%%%%%%%
The discrepancy between the present numerical result and
the CVM analysis suggests that
just like the above mentioned case of $K=K_{G3}$,
the fluctuations beyond a hexagon cluster are important.
As a matter of fact, the preceding 
full-diagonalization analysis for $L\le 6$ \cite{Bowick97} could not even detect
a sign of singularity around $K=K_{P3}$, and the authors speculated that
the singularity would not be captured correctly until $L=8$.
In other words, we need to consider long-length-scale fluctuations exceeding $L=8$,
at least, in order to capture the characteristics of this transition.
In this sense, the crumpling transitions in the $K<0$ regime
are much more subtle than that of $K>0$ studied previously in Refs.
\cite{Bowick96,Nishiyama04}.

%%%%%%%%%%%%%%%%%%%%%%
Moreover, as shown in Fig. \ref{figure9}, the behavior of the excess free energy
$f_{\leftarrow} - f_{\rightarrow}$  is asymmetric with respect to the
transition point $K=-0.76(1)$;
namely, the absolute value of the excess free energy is enhanced in the $K<-0.76$ side,
whereas in the other size $K>-0.76$, it remains suppressed.
(Note that as for $K \sim K_{G3}$, for both sides of the transition,
the excess free energy behaves quite similarly (symmetrically) as shown in Fig. \ref{figure7}.)
A large amount of the excess free energy reflects an existence of strong metastability.
In fact, as shown in Fig. \ref{figure4}, 
the order parameters exhibit notable hysteresis effects in the regime $K<-0.76$.
That is the reason why a symptom of the transition implied by the hysteresis curve
deviates significantly from the true transition point
determined by the coexistence condition (zero excess free energy) 
$f_\leftarrow - f_\rightarrow=0$.
We suspect that the hexagon-cluster-approximation CVM estimate $K_{P3}=-0.852$,
which is slightly smaller than ours $K_{P3}=-0.76(1)$,
could also be biased by this asymmetry.
Nevertheless, as mentioned in the Monte Carlo simulation \cite{Pujol04},
such a metastability, reminiscent of that of the spin glasses, is quite peculiar,
and further details have be clarified
so as to confirm the present simulation result.

In order to estimate the latent heat,
we plotted the residual internal energy $E_\leftarrow-E_\rightarrow$ 
in Fig. \ref{figure10}.
From the residual internal energy at the transition point,
we estimate the latent heat as $Q=0.03(2)$.
As noted above, 
such a finite latent-heat release is due to the fluctuation effect
beyond a meanfield level.
It is to be noted however,
in the present case of $K=K_{P3}$, the fluctuation effect leads to enhancing
the latent-heat release
rather than smearing it out as in the above mentioned case of $K=K_{G3}$.

We notice that the amount of the latent heat is considerably suppressed.
(Note that the latent heat at $K=K_{C3}(>0)$ is much larger 
$Q=0.365(5)$ \cite{Nishiyama04}. )
Our simulation result suggests that the
long-length-scale fluctuations beyond the meanfield treatment 
can drive the transition to a {\em weak}-first-order type.
As mentioned in Introduction, the singularity of the crumpling transition
for (realistic) membranes is arousing controversies.
In this respect, the present result cautions us not to
rule out the possibility of weak-first-order type very easily
so as to avoid misidentifying it 
with a continuous one of an unknown universality class.
Nevertheless, we suspect that the transition in the $K<0$ side
is relevant to the crumpling transition of (realistic) tethered membranes
rather than
the transition in the other side $K<0$,
whose discontinuous character is too pronounced.

According to the analytical theory \cite{Bowick97}, 
the crumpling transition at $K=K_{P3}$
is equivalent to that of the two-dimensional (planar) folding
\cite{Kantor90,DiFrancesco94a,DiFrancesco94b,Cirillo96b},
which exhibits a continuous transition in $K<0$ \cite{Cirillo96b}.
Hence,
provided that this equivalence holds, the crumpling transition at $K=K_{P3}$
should be continuous as well.
This mapping is based on the idea that for large $|K|$,
the complete and acute folds are dominant, and
within 
this restricted configuration space, the discrete folding is equivalent to 
the planar folding.
%%%%%%%%%%%%%%%%%%%%
%%%%%%%%%%%%%%%%%%
However, our simulation result does not support this claim,
and rather, it suggests that the restricted-configuration picture
would not be validated.
Possibly, a slight violation of such a restriction alters
the transition singularity into a {\em weak}-first-order type.
In other words, in reality,
various types of fluctuations are promoted around $K=K_{P3}$,
and
such fluctuations give rise to a finite latent-heat release.

\section{\label{section4}Summary and discussions}

We investigated the singularities of the crumpling transitions
of the discrete-folding model in the negative-bending-rigidity regime $K<0$.
We adopted the density-matrix renormalization group to diagonalize
the transfer matrix for the strip widths up to $L=26$;
the system sizes treated here are substantially larger
than those tractable with
the conventional full-diagonalization scheme $L \le 6$ \cite{Bowick97}.
Actually, in Ref. \cite{Bowick97}, 
it was speculated that the singularity at $K=K_{P3}$
would not be captured correctly until $L=8$.
Hence, it is significant to consider long-length-scale undulations
exceeding $L=8$.
In order to check the reliability of the present simulation,
we calculated the distribution of the density-matrix
eigenvalues (statistical weights) for $L=15$ and $K=-0.2$.
From the data, we found that the truncation error
due to the renormalization group (real-space decimation) procedure is of the order
$\sim 5 \times 10^{-4}$ for $m=15$ (number of bases remained for a renormalized block).
This result is encouraging in the sense that the truncation error is smaller than
the finite-size corrections, and in practice, almost negligible.

\begin{figure}
\includegraphics{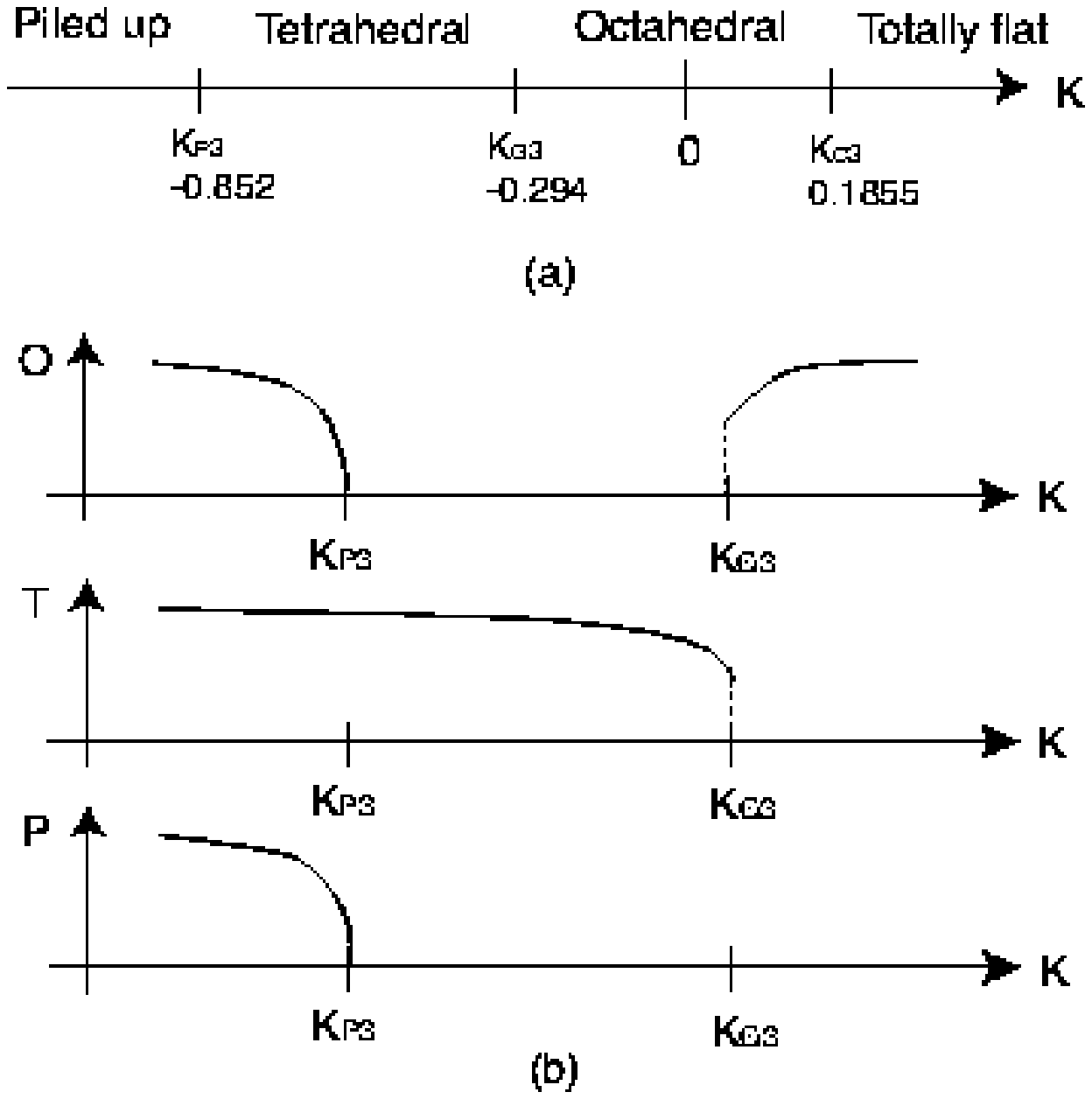}%
\caption{\label{figure1}
(a) A phase diagram of the discrete folding model
determined with the cluster-variation method (CVM) \cite{Bowick97}.
Various phases appear as the bending rigidity $K$ changes.
(b) A schematic drawing of the behaviors of the order parameters, 
Eqs. (\ref{order1})-(\ref{order2}), determined with CVM \cite{Bowick97};
this result is to be compared with our first-principles-simulation data 
as shown in Fig. \ref{figure4}.
Note that
according to CVM, the transition at $K_{P3}$ ($K_{G3}$) should be
a continuous (discontinuous) one. 
}
\end{figure}

\begin{figure}
\includegraphics{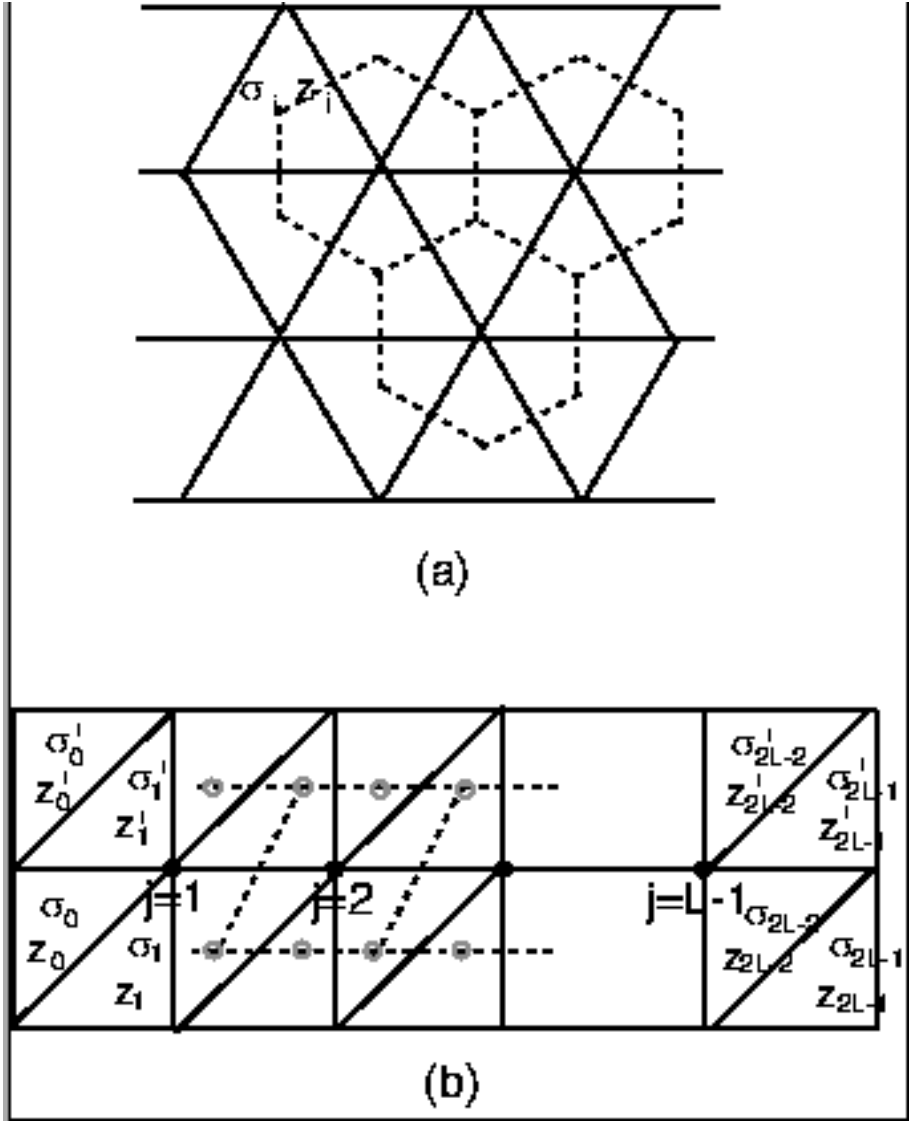}%
\caption{\label{figure2}
(a) We consider a discrete folding of the triangular lattice
embedded in the three-dimensional space.
In order to specify the fold angle, we place two types of
Ising variables such as $z_i$ and $\sigma_i$ at each triangle
rather than at each joint (gauge rule \cite{Bowick95}).
Hence, hereafter, we consider a spin model 
defined on the dual (hexagonal) lattice.
(b) A construction of the transfer matrix.
The row-to-row statistical weight yields the transfer-matrix element.
The explicit formula is given by Eq. (\ref{transfer_matrix}).
The transfer matrix is diagonalized with the density-matrix renormalization
group; see Fig. \ref{figure3}.
}
\end{figure}

\begin{figure}
\includegraphics{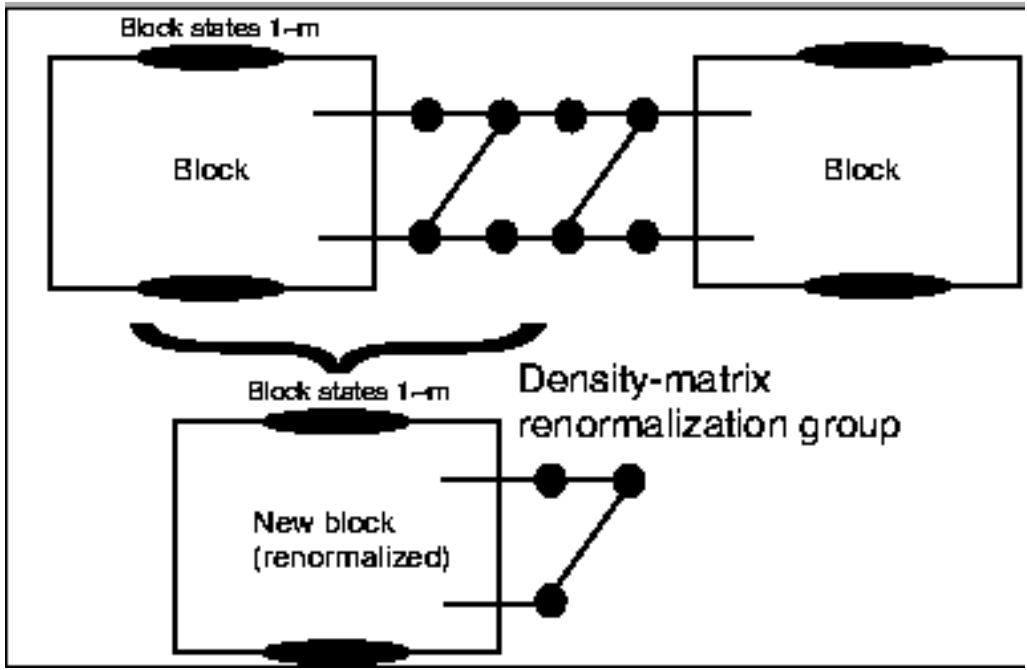}%
\caption{\label{figure3}
Schematic drawing of the density-matrix renormalization group (DMRG) 
procedure.
From the drawing, we see that through one operation of DMRG,
a ``block'' and the adjacent sites (hexagon) are 
renormalized altogether into a new renormalized ``block.''
At this time, the number of block states is retained within $m$;
see text for details.
In this manner,
we can diagonalize a large-scale transfer matrix
through sequential applications of DMRG.
}
\end{figure}

\begin{figure}
\includegraphics{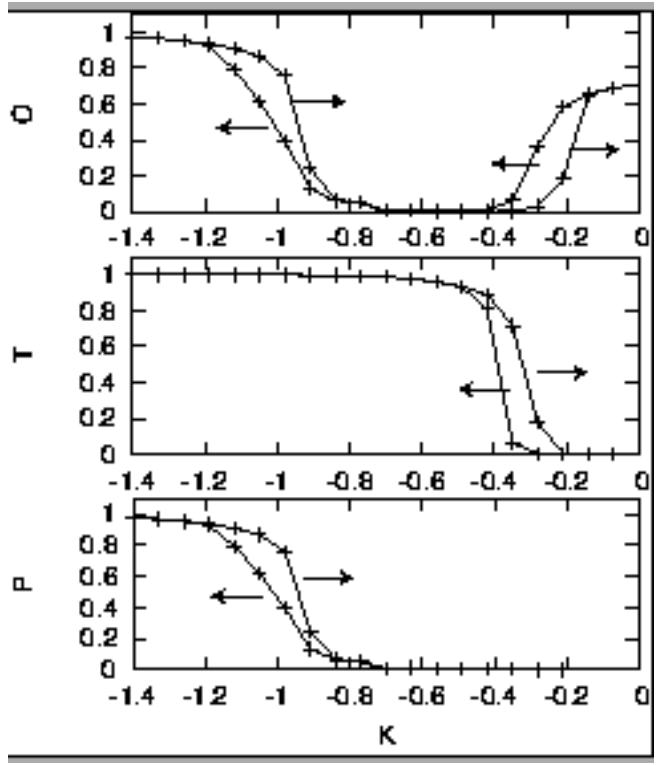}%
\caption{\label{figure4}
The order parameters, Eqs. (\ref{order1})-(\ref{order3}), are calculated with the
density-matrix renormalization group for the system size $L=14$ and the number of states
kept for a block $m=12$.
The arrows indicate the directions of the parameter sweeps.
We notice that the behaviors are similar to those obtained
with the CVM analysis (see Fig. \ref{figure1}).
However, contrary to the prediction,
the simulation data exhibit clear evidences of the hysteresis effects for both transitions,
implying that these transitions are both discontinuous.
}
\end{figure}

\begin{figure}
\includegraphics{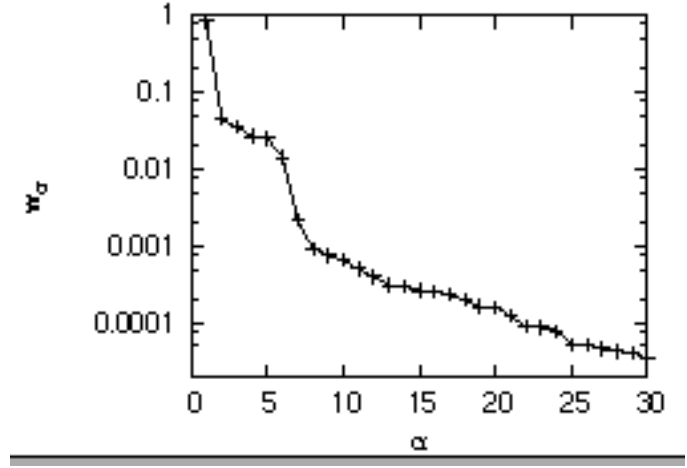}%
\caption{\label{figure5}
Distribution of the eigenvalues (statistical weights)
$\{ w_\alpha \}$ ($\alpha$, integer index) of the (local) density matrix 
\cite{White92,White93,Nishino95} is shown.
The simulation parameters are chosen from 
the bending rigidity $K=-0.2$, the number of states kept for a block  $m=15$,
and the system size $L=15$.
We see that $w_\alpha$ drops very rapidly for large $\alpha$.
Because we discard the irrelevant states 
$\alpha>m$, 
the weight $w_{\alpha}$ at $\alpha=m$ yields an indicator of the 
truncation error.
We attain a precision of the order $5 \times 10^{-4}$ under setting $m=15$ for instance.
%For instance,
%the states up to only $\alpha=15$ cover the relevant (significant) bases with
%appreciable statistical weights $w_\alpha > 3\cdot 10^{-4}$;
%the discarded weights ($\alpha>m$)
%indicate an amount of the truncation error.
%In this manner, only $m$ selected (relevant) bases are retained for a
%$renormalized ``block'' as is shown in Fig. \ref{figure3}.
}
\end{figure}

\begin{figure}
\includegraphics{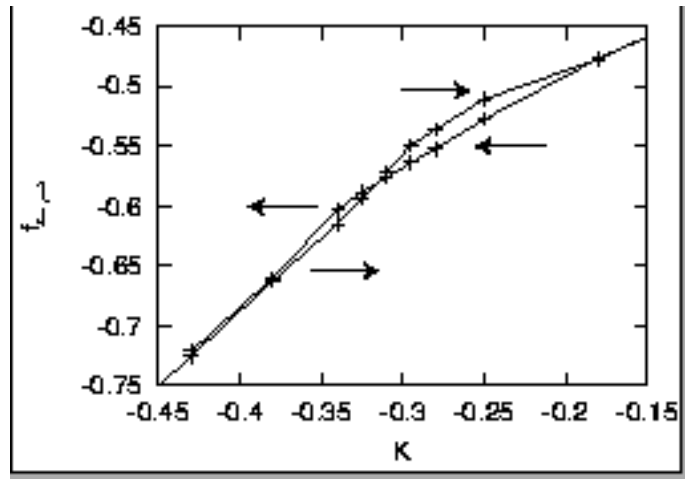}%
\caption{\label{figure6}
Free energy per triangle $f_{\rightarrow,\leftarrow}$
is plotted for $L=20$ and $m=15$.
These subscripts, $\rightarrow$ and $\leftarrow$,
specify the directions of the parameter sweeps,
namely, the ascending and descending directions, respectively.
We see that at $K \approx -0.3$, a first-order phase transition occurs.
}
\end{figure}

\begin{figure}
\includegraphics{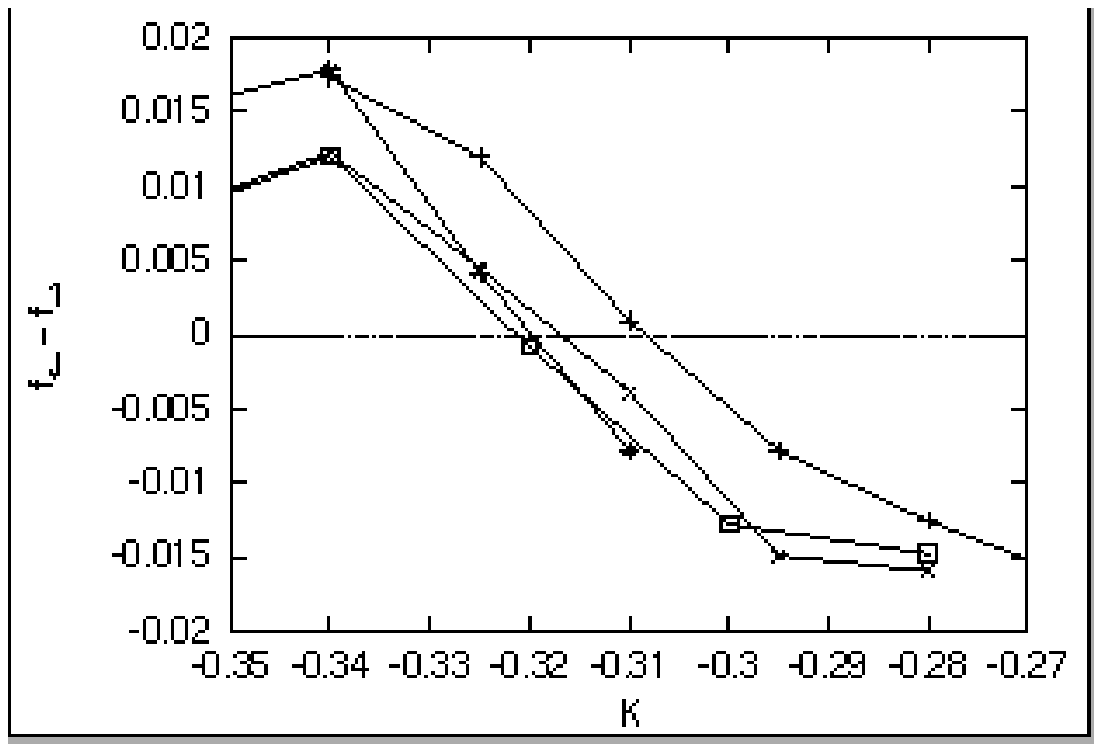}%
\caption{\label{figure7}
Excess free energy $f_\leftarrow-f_\rightarrow$ is plotted for 
($+$) $L=18$ and $m=15$,
($\times$) $L=20$ and $m=15$,
($\ast$) $L=20$ and $m=20$,
and
($\Box$) $L=22$ and $m=15$.
From the coexistence criterion $f_\leftarrow-f_\rightarrow=0$, 
we determined the transition point as $K=-0.32(1)$.
}
\end{figure}

\begin{figure}
\includegraphics{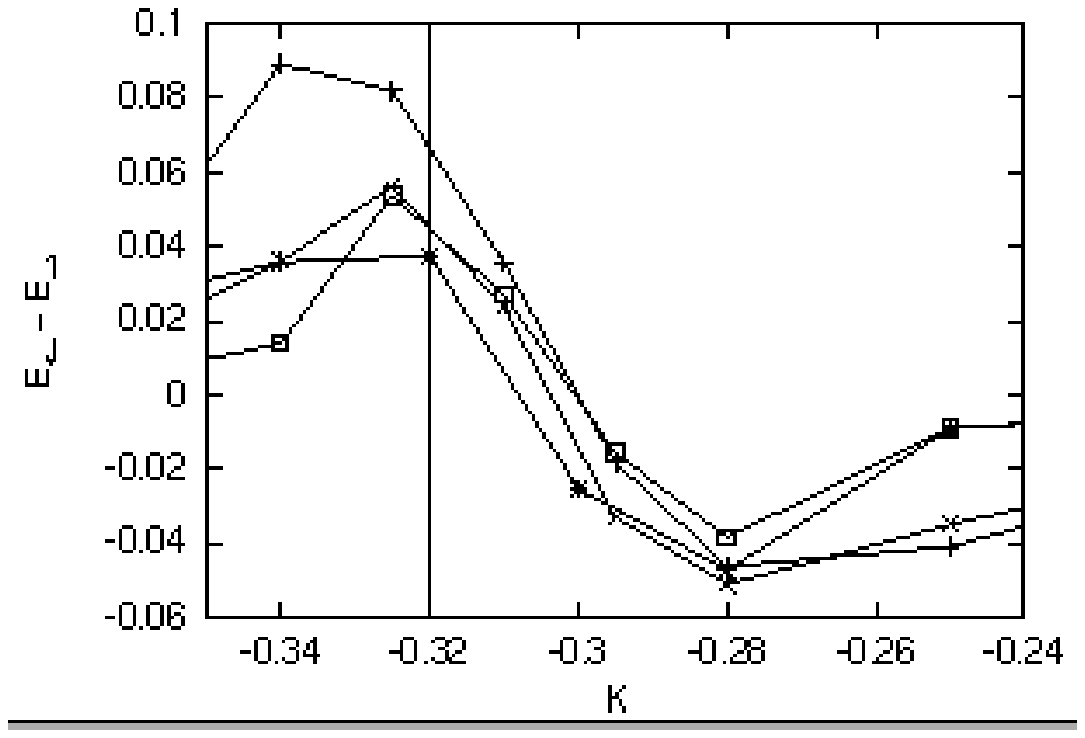}%
\caption{\label{figure8}
Residual internal energy $E_\leftarrow-E_\rightarrow$ per triangle is plotted for
($+$) $L=18$ and $m=15$,
($\times$) $L=20$ and $m=15$,
($\ast$) $L=22$ and $m=15$,
and
($\Box$) $L=24$ and $m=15$.
From the residual internal energy at the transition point,
we estimate the latent heat as $Q=0.04(2)$.
}
\end{figure}

\begin{figure}
\includegraphics{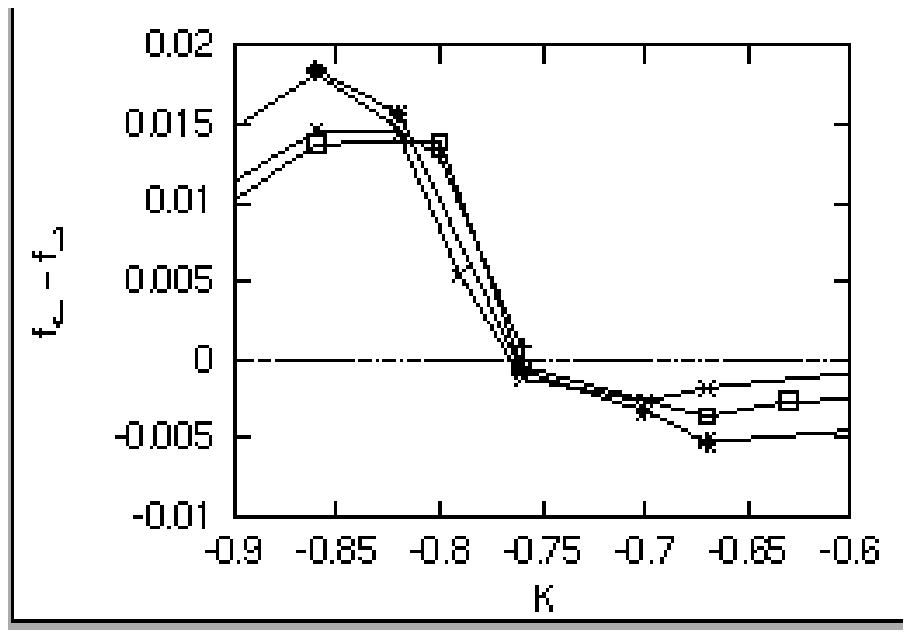}%
\caption{\label{figure9}
Excess free energy $f_\leftarrow-f_\rightarrow$ is plotted for 
($+$) $L=18$ and $m=20$,
($\times$) $L=20$ and $m=15$,
($\ast$) $L=22$ and $m=15$,
and
($\Box$) $L=22$ and $m=20$.
From the coexistence criterion $f_\leftarrow-f_\rightarrow=0$, 
we determined the transition point as $K=-0.76(1)$.
}
\end{figure}

\begin{figure}
\includegraphics{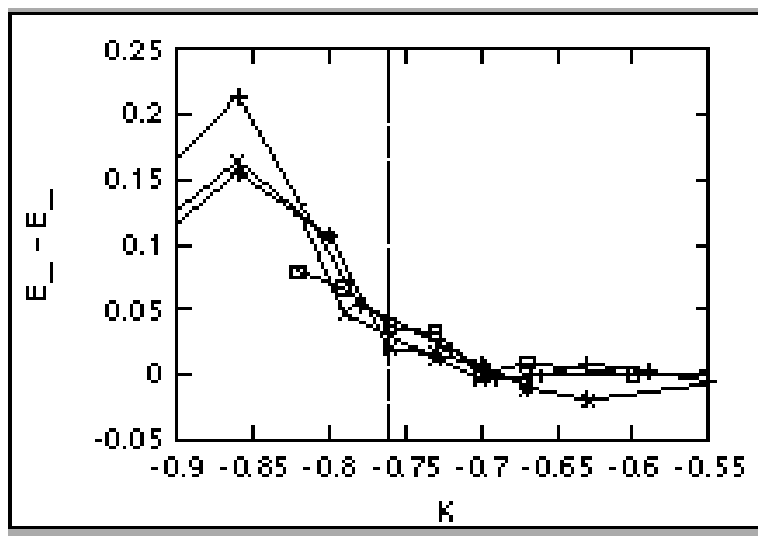}%
\caption{\label{figure10}
Residual internal energy $E_\leftarrow-E_\rightarrow$ per triangle is plotted for
($+$) $L=19$ and $m=15$,
($\times$) $L=20$ and $m=15$,
($\ast$) $L=22$ and $m=20$,
and
($\Box$) $L=26$ and $m=15$.
From the residual internal energy at the transition point,
we estimate the latent heat as $Q=0.03(2)$.
}
\end{figure}

As a preliminary survey,
we calculated the order parameters, Eqs. (\ref{order1})-(\ref{order3}), 
for a wide range of $K(<0)$.
The overall behaviors support the CVM scenario
as depicted in Fig. \ref{figure1};
namely, there appear
two crumpling transitions around $K \approx -0.9$ and $-0.3$, which separate 
the piled-up, tetragonal and octahedral phases.
However, on closer inspection, we observe clear hysteresis effects
for both transitions.
That is, contrary to the CVM scenario, both singularities should be discontinuous.

Performing large-scale simulations, we determined the transition points as $K=-0.76(1)$ and $-0.32(1)$.
At respective transition points, we estimated the amounts of the latent heat as
$Q=0.03(2)$ and $0.04(2)$.
Comparing the present results with the CVM estimates with the 
single-hexagon-cluster approximation, namely,
$Q=0$ ($K_{P3}=-0.852$) and $0.14$ ($K_{G3}=-0.294$),
we notice that the discrepancies between them are rather conspicuous.
The present simulation results suggest that the fluctuations beyond 
a hexagon cluster are significant.
According to Ref. \cite{Bowick97},
the transition at $K=K_{P3}$ should be a continuous one, provided that 
complete and acute folds are dominant around the transition point;
within this restricted configuration space, the discrete folding is
equivalent to the planar (two-dimensional) folding, which exhibits a
continuous transition in $K<0$.
However, our simulation does not support this claim,
because the transition turns out to be of a discontinuous character.

In an earlier paper \cite{Nishiyama04}, 
we studied the crumpling transition in $K>0$,
and reported that it belongs to a first-order phase transition with the latent heat $Q=0.365(5)$.
Consequently, it turns out that all crumpling transitions 
of the discrete-folding model are discontinuous.
However, it is notable that the amounts of the latent heat in $K<0$
are much smaller than that of the other side $K>0$.
In other words, the singularities of the crumpling transitions in $K<0$ are
all
{\em weak}-first-order phase transitions.
As mentioned in Introduction, on the contrary,
a majority of Monte Carlo simulation for (realistic) tethered membranes
has reported that the crumpling transition should be a second-order phase transition
\cite{Kantor86,Kantor87,Baig89,Ambjorn89,Renken96,Harnish96,%
Baig94,Bowick96b,Wheater93,Wheater96,David88,Doussal92,Espriu96}.
However, as noted in Refs. \cite{Kantor87b,Kantor87c},
it is not quite obvious to rule out the possibility of weak 
first-order transition in practice.
Hence, it would be desirable to set forward the present analysis to 
other discretized-membrane models,
such as the square-diagonal-lattice folding and its variants
\cite{DiFrancesco98a,DiFrancesco98b,Cirillo00},
in order to see whether the crumpling transitions are still of discontinuous
characters
(weak first order);
the discretized-membrane model admits 
detailed  analysis on the character of the crumpling transition.
This problem will be addressed in future study.

\begin{acknowledgments}
This work is supported by a Grant-in-Aid for
Young Scientists
(No. 15740238) from Monbukaagakusho, Japan.
\end{acknowledgments}

% Create the reference section using BibTeX:

\end{document}